\documentclass[journal]{IEEEtran}

\usepackage{xcolor,soul,framed} 

\usepackage[pdftex]{graphicx}
\graphicspath{{../pdf/}{../jpeg/}}
\DeclareGraphicsExtensions{.pdf,.jpeg,.png}

\usepackage[cmex10]{amsmath}
\usepackage{siunitx}
\usepackage{graphicx}
\usepackage{multirow}
\usepackage{multicol}
\usepackage{amssymb} 
\usepackage{amsmath}
\usepackage{mathtools}
\usepackage{cite}
\usepackage{siunitx}
\usepackage{cuted}
\usepackage{flushend}
\usepackage{multibib}
\usepackage{relsize}
\usepackage{stix}
\usepackage{balance}
\usepackage{array, makecell}

\DeclarePairedDelimiterXPP\BigOSI[2]{\mathcal{O}}{(}{)}{}{\SI{#1}{#2}}

\begin{document}
\bstctlcite{IEEEexample:BSTcontrol}
    \title{Random Design Variations of Hollow-core Anti-resonant Fibers: A Monte-Carlo Study}
  \author{
           Michael Petry,~\IEEEmembership{Student Member,~IEEE}, Md. Selim Habib,~\IEEEmembership{Senior Member,~IEEE,} ~\IEEEmembership{Member, OSA}

  \thanks{Manuscript received February $\times\times$, 2022; $\times\times$, 2022; accepted $\times\times$, 2022. Date of publication $\times\times$,
2022; date of current version $\times\times$, 2022. 
(\emph{Corresponding author:
Md. Selim Habib.})}
   \thanks{M. Petry and M. Selim Habib are with the Department of Electrical and Computer Engineering, Florida Polytechnic University, FL-33805, USA (e-mail: mhabib@floridapoly.edu).}
   }

\markboth{IEEE Journal of Lightwave Technology, VOL.~XXX, NO.~XXX, May~2022}
{Habib \MakeLowercase{\textit{et al.}}:XXX}

\maketitle


\begin{abstract}
Hollow-core anti-resonant fibers (HC-ARFs) have earned great attention in the fiber optics community due to their remarkable light-guiding properties and broad application spectrum. 
Particularly nested HC-ARFs have recently reached competitiveness to standard fibers and even outperform them in certain categories. Key to their success is a precisely fine-tuned geometry, which inherently leaves optical characteristics highly susceptible to minimal structural deviations. When fabricating fibers, these come into play and manifest themselves in various imperfections to the geometry, ultimately worsening the fiber performance. In this article, for the first time to the best of our knowledge, these imperfections have been statistically modeled and analyzed on their impact to the propagation loss in a Monte-Carlo fashioned simulation. We have considered randomly varying outer and nested tube wall thicknesses as well as random tube angle offsets. We found that the loss increase due to perturbed tube angles dominates that of varying tube thicknesses by approximately an order of magnitude for FM and two orders of magnitude for HOM propagation at a wavelength of \SI{1.55}{\micro\meter}. Moreover, the higher-order-mode-extinction-ratio (HOMER) is proportional to the intensity of structural variations, indicating an increase in the `single-modeness' of a fabricated fiber. Furthermore, a bend condition worsens the loss contribution of both effects applied jointly dramatically to a value of $+50\%$ at a bend radius of \SI{4}{\centi\meter} compared to $+5\%$ for a straight fiber. We believe that our work helps to predict the performance of realistic HC-ARFs after fabrication.
\end{abstract}


\IEEEpeerreviewmaketitle

\section{Introduction}
\IEEEPARstart{L}{ight} guidance in air-core hollow-core anti-resonant fibers (HC-ARFs) (also called negative curvature HC-ARFs or tubular HC-ARFs) finds immense interest in the fiber optics society due to their superior and unique optical properties~\cite{pryamikov2011demonstration,belardi2014hollow_nested,poletti2014nested,habib2019single,sakr2019ultrawide,sakr2020interband,bradley2018record,bradley2019antiresonant,debord2013hypocycloid,debord2017ultralow,yu2016negative,yu2012low,van2016modal,gao2018hollow,jasion2020hollow,habib2016low}. HC-ARFs offer low latency, low non-linearity, extremely low power overlap with silica glass which allows high damage threshold and reduced material attenuation, and low anomalous dispersion over a wide transmission range that can not be attained in silica-based solid-core fibers~\cite{poletti2014nested}. Such remarkable optical properties of HC-ARFs find numerous applications which includes but is not limited to high power delivery~\cite{michieletto2016hollow,gebhardt2017nonlinear}, gas-based nonlinear optics~\cite{travers2011ultrafast,russell2014hollow,adamu2019deep,adamu2020noise,habib2017soliton,markos2017hybrid,wang2020high}, extreme UV light generation~\cite{habib2019extreme,habib2018multi}, non linear microendoscopy~\cite{kudlinski2020double}, mid-IR transmission~\cite{kolyadin2013light,urich2013flexible}, optofluidic \cite{hao2018optimized}, and terahertz applications~\cite{anthony2011thz,hasanuzzaman2015low,sultanan2020exploring}. Recently, HC-ARFs are also used in short-reach data transmission~\cite{sakr2020interband,sakr2019ultrawide}, next-generation scientific instruments and polarization purity~\cite{Taranta2020Exceptional}.

The HC-ARF guides light in the air-core based on inhibited coupling (IC) between the core-guided modes and the continuum of modes of the cladding \cite{pearce2007models}, including also anti-resonant effect~\cite{poletti2014nested}. The concept of IC guiding mechanism was first proposed by Couny \emph{et. al.,}~\cite{couny2007generation}. The IC guiding mechanism can be explained by the high degree of the transverse-field mismatch between the core and cladding modes (CMs). In HC-ARFs, the IC between the core mode and CMs can be enhanced dramatically by using a negative curvature core contour~\cite{debord2013hypocycloid}, carefully engineering the cladding structure and choosing a proper number of cladding tubes~\cite{habib2019single,habib2021impact}. Recently, it has been reported that the number of tubes in the cladding structure plays one of the critical factors in significantly reducing the propagation loss and maintaining effectively single-mode operation \cite{habib2021impact,habib2019single}. The enhanced IC guiding of HC-ARFs offers a much wider transmission window, low dispersion, and lower loss compared to hollow-core photonic bandgap fibers (HC-PBGFs) in which light guides based on the photonic bandgap effect~\cite{cregan1999single,amezcua2008control}. 

Recently, several HC-ARF designs based on a 'negative curvature' core surround have been studied and demonstrated towards realizing low loss transmission ~\cite{pryamikov2011demonstration,belardi2014hollow_nested,poletti2014nested,habib2019single,debord2013hypocycloid,debord2017ultralow, bradley2018record,sakr2019ultrawide,sakr2020interband,yu2016negative,bradley2019antiresonant}. Most of the designs rely on nested HC-ARFs in which nested anti-resonant tubes are used in the cladding \cite{belardi2014hollow_nested,poletti2014nested,habib2015low}. Adding nested tubes drastically enhances the IC coupling between the core and cladding modes and thus significantly reduces propagation loss and macro bend loss \cite{belardi2014hollow_nested,poletti2014nested}. The number of cladding tubes plays an important role while designing ultra-low loss fibers with effectively single-mode operation. Most of the earlier proposed nested fiber geometries used either eight \cite{belardi2014hollow_nested} or six tubes \cite{poletti2014nested,habib2015low}. In 2019, Habib \it{et al.}\rm, \cite{habib2019single} thoroughly investigated the impact of the number of cladding tubes on the propagation loss and single-modeness of the fiber, and it was found that five-tube HC-ARFs show a lower loss and effectively single-mode operation compared to other tubes. In 2020, a propagation loss of \SI{0.28}{dB/km} from \SI{1510}{\nano\meter} -- \SI{1600}{\nano\meter} had been demonstrated with a six-tube nested HC-ARF design \cite{jasion2020hollow}. Most recently, a five-tube HC-ARF has been fabricated with a record low loss of \SI{0.22}{dB/km} at \SI{1300}{\nano\meter} and \SI{1625}{\nano\meter} \cite{sakr2021hollow}. Recently, a five-tube anisotropic HC-ARF design with a propagation loss of 0.11 dB/km has been proposed  and the impact of nested tubes on the propagation loss are studied. However, in all previous studies, the effect of random structural perturbations to the fiber geometry such as random variations in the tube wall thicknesses and random tube gap separations were not considered. These effects are typically introduced in the fiber fabrication process and might impact the overall loss and single-mode propagation performance.

This work aims to investigate the impact of random structural perturbations on the overall loss performance of nested HC-ARFs and derive a connection between fabrication metrics and the expected fiber performance. With multiple imperfections being present after fabrication, we have selected the two most prominent imperfections, namely random tube thicknesses as well as random tube angle offsets for thorough analysis. By interpreting these effects using statistical models and evaluating them independently and jointly, we provide insights into their impact on the propagation loss under multiple operating conditions. 
The article is organized as follows: Section \ref{section:fibergeoms} reviews the typical nested HC-ARF geometry and introduces two types of random design variations it suffers from after fabrication. Section \ref{section:num_results} optimizes those geometries for the lowest propagation loss and studies its susceptibility to the imperfections introduced prior by performing numerous simulations in a Monte-Carlo fashion. Both single and higher-order mode propagation is considered for a straight fiber configuration, whereas the focus is set on single-mode propagation when introducing a bend condition as well as changing the operational wavelength. Moreover, an outlook on further studies is given. Section \ref{section:conclusion} summarizes this article.

\section{Introduction to Random Design Variations}
\label{section:fibergeoms}
\begin{figure}[t]
  \begin{center}
  \includegraphics[width=3in]{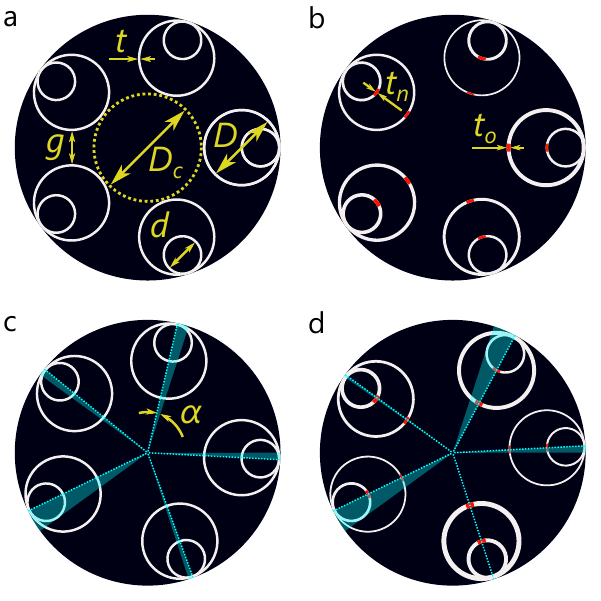}\\
  \caption{Random design variations analyzed in this article illustrated for a five-tube HC-ARF. (a) Ideal (symmetric) nested HC-ARF design with five circular tubes; (b) allowing individual outer and nested tube thicknesses, $t_o$ and $t_n$, respectively; (c) allowing individual angle offsets for tubes; (d) geometry features of (b) and (c) combined concurrently. The thickness of tubes drawn is exaggerated and not to scale for improved visibility. The design parameters are similar to \cite{sakr2019ultrawide,sakr2020interband}.}\label{fig:fig_1}
  \end{center}
\end{figure}

The different types of random design variations analyzed in this work are displayed in Fig. \ref{fig:fig_1}. For the purpose of illustration, we have chosen to demonstrate these effects for a five-tube nested HC-ARF geometry since we will only focus on this structure in our random design analysis. 
However, the techniques introduced here are equally applicable for any other number of tubes. 
The \textit{ideal} geometry of a five-tube HC-ARF is shown in Fig. \ref{fig:fig_1}a. In this work, we have chosen a core diameter of $D_\text{c}=\SI{35}{\micro\meter}$, similar to \cite{habib2021impact}. The outer diameter $D$ and nested tube diameter $d$ depend on the core diameter, number of cladding tubes $N$, and tube gap separation $g$, which will be determined in the optimization part in subsection \ref{chapter:geom_optim}. The nested tube ratio $d/D$ is kept fixed to 0.5 since this ratio effectively provides single-mode operation \cite{habib2019single,habib2021impact}. Lastly, the tube wall thickness, usually in the range of multiple \SI{100}{\nano\meter}'s will also be determined in the optimization step due to its strong impact on the propagation loss. The ideal structure is characterized by constant fiber parameters. However, when fabricating HC-ARFs with sub-$\SI{}{\micro\meter}$ geometry features in practice, this assumption no longer holds~\cite{sakr2019ultrawide}. With many imperfections coming into play, two distinct imperfections have been identified to considerably impact the geometry, which will be thoroughly addressed in this work. For example, maintaining the same wall thickness for both tubes is typically challenging when fabricating $<\SI{400}{\nano\meter}$ thick tubes \cite{sakr2020interband}. We assume the wall thicknesses of the cladding tubes to deviate up to $\pm 5\%$ from the target thickness. Also, recent studies show that the mean nested tube thickness $t_n$ varies around $\pm10\%$ of the outer tube thickness $t_o$ \cite{habib2021impact,sakr2020interband,sakr2021hollow}. 
Moreover, not only do the tubes' thicknesses vary, but also the absolute tube positions itself as displayed in Fig. \ref{fig:fig_1}c. This manifests itself in an individual offset to the tube angles, denoted by $\alpha_1$... $\alpha_5$, each introducing a tube rotation around the geometrical center of the fiber. The angular deviation of the tubes from their original axes is illustrated by a light-blue shaded area, with a dotted line representing the original axis. Another way to look at this phenomenon is from the perspective of varying gap separations between adjacent tubes. Of course, those quantities depend on each other and can be converted interchangeably, which is shown in Eq. \eqref{formula:gapangle}.
The gap separation $g_{12}$ between adjacent tubes one and two depends on the core diameter $D_\text{c}$, outer-tube diameter $D$\cite{habib2021impact}, outer-tube thickness $t$ (assuming equal thicknesses), and the respective angle offsets $\alpha_1$ and $\alpha_2$:
\begin{subequations}
\renewcommand{\theequation}{\theparentequation.\arabic{equation}}
\begin{gather}
g_{12} = \sqrt{2(\frac{D_c}{2}+\frac{D}{2}+t)^2\left[1-\cos{\left(\frac{2\pi}{N}+{\alpha}_2-{\alpha}_1\right)}\right]} - D - \frac{t}{2}\label{formula:gapangle} 
\intertext{By geometrical dependency, the outer tube diameter $D$ can also be calculated from different parameters, namely the number of tubes $N$ and the mean gap separation $g_\text{mean}$ according to \cite{wei2017negative}:}
D = \frac{\frac{D_c}{2}sin(\frac{\pi}{N})-\frac{g_\text{mean}}{2}-t(1-sin(\frac{\pi}{N}))}{1-sin(\frac{\pi}{N})}
\end{gather}
\end{subequations}
Due to advantages in the handling of angle offsets, e.g., the statistical independence of $\alpha_1$ to $\alpha_5$ (contrary to $g_1$ to $g_5$) and the intuitive connection to the manufacturing quality metrics, we will pursue the angle offset perspective for the course of this article. Conversion methods, e.g., from angle offset standard deviation to gap separation standard deviation, are left subject to further research. The influence of both effects (wall thickness variations and angle offsets) on the propagation loss was investigated independently as shown in Fig. \ref{fig:fig_1}(b-c). Next, the impact on the propagation loss was studied considering both effects simultaneously as shown in Fig. \ref{fig:fig_1}d.

To support the effects described prior, we implemented additional degrees of freedom into the simulation model, namely individual nested and outer tube thicknesses $t_{o,1}$ ... $t_{o,5}$ and $t_{n,1}$ ... $t_{n,5}$, respectively, as well as individual tube angle offsets $\alpha_1$ ... $\alpha_5$.
In the course of this article, we will incorporate statistical properties based on practically feasible geometrical characteristics.


\section{Numerical Results and Discussion}
\label{section:num_results}
All simulations were performed using finite-element analysis based on \textsc{Comsol}$^{\circledR}$ software in combination with MATLAB-Livelink. To accurately calculate the modal properties of the fiber, we placed a perfectly-matched layer (PML) outside the fiber domain. The simulation environment and the modeling parameters, i.e., the mesh element size in the air and silica regions and PML boundary conditions were configured similarly to \cite{poletti2014nested,habib2019single,habib2021impact}. However, since we introduce spatial random structural variations across the HC-ARF geometry, symmetry inherent under ideal conditions is voided. This makes the well-used practice of exploiting this symmetry by solving only half of the structure using proper boundary conditions no longer feasible. We, therefore, solve the whole domain of the fiber, which leads to an increase in simulation time and memory consumption. For this paper, the primary HC-ARF characteristic we specifically focus on is transmission loss. Therefore, we consider multiple types of losses, which constitute of confinement/leakage loss (CL), surface scattering loss (SSL), and macro bend loss (if applicable). Because of the insignificant power overlap with silica walls $<10^{-4}$, effective material loss (EML) has been neglected and is therefore not included in the calculations. 

\subsection{Geometry Optimization}
\label{chapter:geom_optim}
A point of reference is required to analyze and compare the effects of random structural variations on the propagation loss to the principal HC-ARF geometry as described in section \ref{section:fibergeoms}. We define this point of reference as the lowest loss exhibited by the so-called ideal geometry. To identify those points, a propagation loss optimization for five-, six-, seven-, and eight-tube HC-ARFs using a two-dimensional sweep over the tube thickness and gap separation has been performed. An appropriate starting point for the tube thickness is chosen so that the operating wavelength $\lambda_m=\SI{1.55}{\micro\meter}$ sits in between the surrounding high loss resonant wavelengths of the fiber as described by \cite{poletti2014nested}: $\lambda_r=\frac{2t}{m}\sqrt{n_g^2-1}$; n is the refractive index of silica and m is the resonance order. This leads to $t=\SI{370}{\nano\meter}$. 

\begin{figure}[t!]
  \begin{center}
  \includegraphics[width=3.4in]{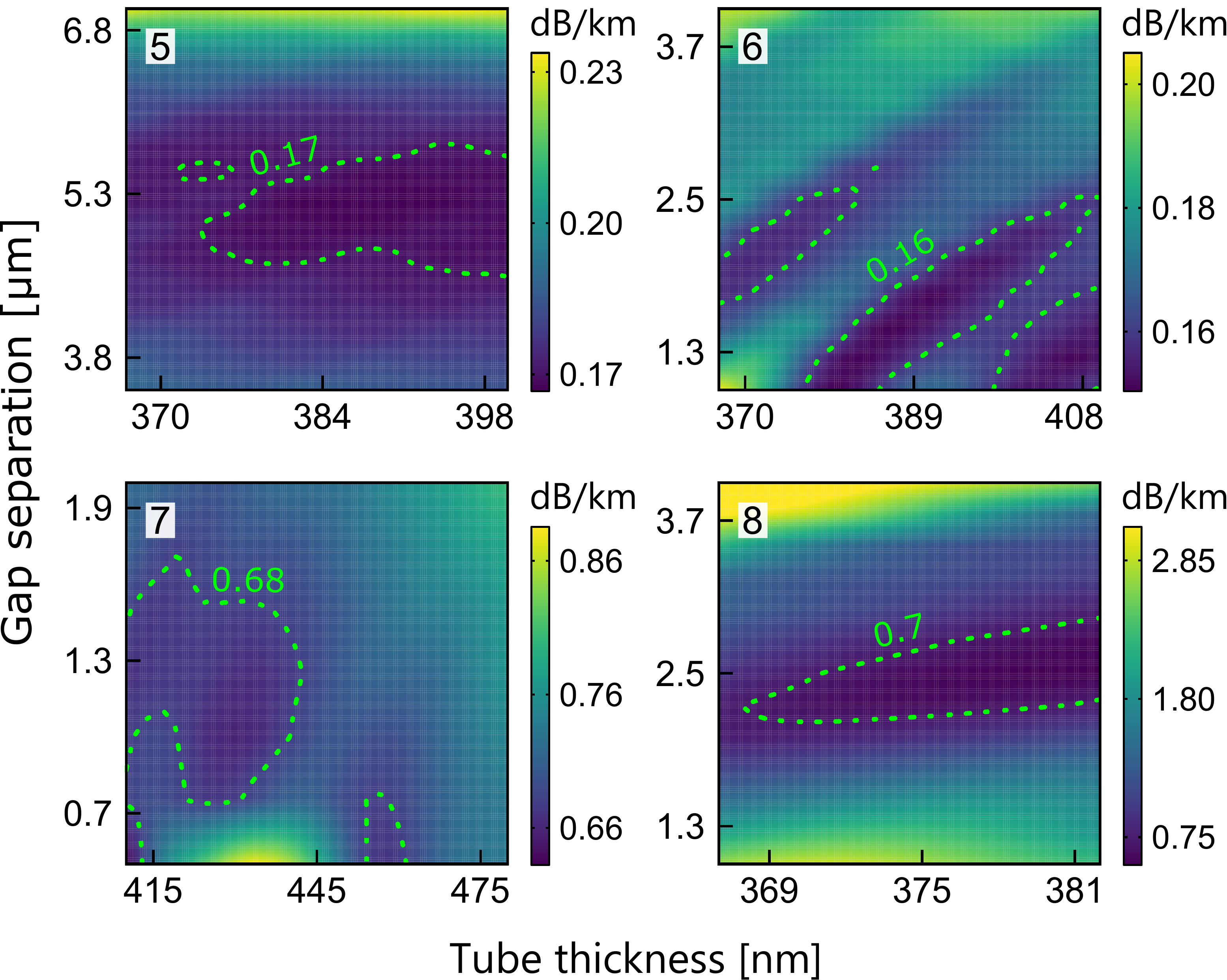}\\
  \caption{Calculated propagation loss of $\text{LP}_\text{01}$-like FM as a function of the tube thickness $t$ and gap separation $g$ for nested HC-ARFs with an ideal geometry in a (5) five-, (6) six-, (7) seven- and (8) eight-tube configuration. The 2D surface plot is generated using a sweep quantization of $\Delta t=\SI{1}{nm}$ and $\Delta g=\SI{0.25}{\micro\meter}$ with linear interpolation.
  Additional design parameters: Core diameter $D_\text{c} = \SI{35}{\micro\meter}$; equal outer and nested tube thickness, $t_o=t_n$; nested/outer tube diameter ratio $d/D=0.5$. All simulations are performed at wavelength $\lambda=\SI{1.55}{\micro\meter}$.}\label{fig:fig_2}
  \end{center}
\end{figure}

The gap separation  $g=\SI{2}{\micro\meter}$ has been chosen based on optimization as seen in \cite{poletti2014nested}. Since the initial conditions are independent of the number of tubes, the searched area has been individually expanded into and around the respective low loss regions to provide insights into the loss behavior around their point of lowest propagation loss. The result is displayed in Fig. \ref{fig:fig_2}. The six-tube structure was found to have the lowest overall propagation loss with values $<\SI{0.16}{dB/km}$, followed closely by the five-tube geometry with values $<\SI{0.17}{dB/km}$. The seven- and eight-tube geometries, on the other hand, start at a fourfold and fivefold loss increase, respectively, making them unsuitable for ultra-low loss transmission within the current geometrical boundaries. The five-tube structure has a significantly wider low loss transmission window than all other geometries and offers the current record-holding lowest loss of \SI{0.11}{dB/km}  given minor modifications, e.g., anisotropic nested tubes \cite{habib2021impact, petryhabib2021}. For those reasons, we expect this fiber to maintain a stable transmission behavior and therefore have chosen the five-tube geometry to be extensively investigated under the influence of random design variations caused by fiber fabrication. The optimum five-tube geometry determined in this simulation has a tube wall thickness of $t=\SI{393}{\nano\meter}$ and a gap separation of $g=\SI{5.25}{\micro\meter}$ and exhibits a minimum loss of \SI{0.17}{dB/km}. This will serve as the reference point for the following investigations.

\subsection{Impact of random design variations to fundamental-mode and higher-order-mode propagation}
\label{section:impact_random_design}
This section aims to study the susceptibility of critical fiber characteristics for fundamental mode (FM) and higher-order-mode (HOM) propagation to random structural variations. 
As mentioned in section \ref{section:fibergeoms}, fabricated fibers are subject to structural imperfections introduced by the manufacturing process. The outer and nested tubes hereby represent a great susceptibility to process variation, whereas two effects, namely a deviation of the tubes' thicknesses and the tube positions itself are believed to pose the major contribution \cite{habib2021impact}. 

In order to represent the process of random perturbations in our calculations, we assume that structural variations are distributed in a zero-mean Gaussian fashion, essentially meaning that small offsets are more common than big offsets. 
Special care must be taken when applying the offsets to the tube angles. To avoid collisions of two adjacent tubes with large, contrary angle offsets, the angle offset distributions have been symmetrically truncated at a cut-off angle of $\alpha_{\text{cut-off}}=\SI{4.5}{\degree}$. Strictly speaking, trimming the tails of a normal distribution decreases the resulting standard deviation of the sampled values since out-of-range values get replaced by smaller, in-range values. In this paper, we always refer to the original standard deviation. To make the model even more realistic, we are assuming a general 10$\%$ increase in the mean thickness of the nested tubes relative to the outer tubes, which has been observed for fabricated fibers due to the manufacturing process. This increase is also reflected in the nested tube thickness standard deviation. However, one could also assume a decreased nested tube thickness \cite{sakr2021hollow}.
\begin{figure}
  \begin{center}
  \includegraphics[width=3.4in]{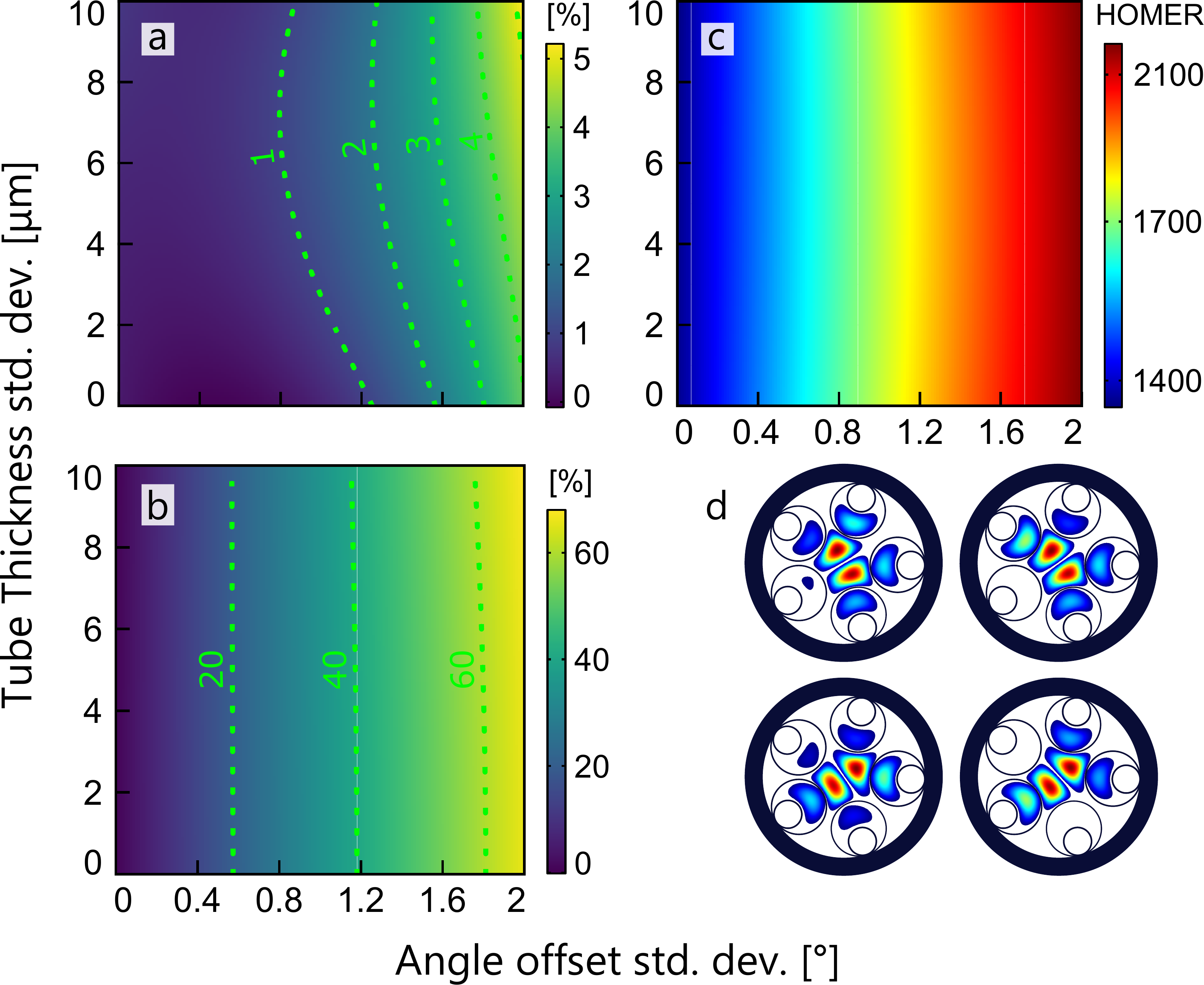}\\
  \caption{Calculated median propagation loss of (a) $\text{LP}_\text{01}$-like FM and (b) $\text{LP}_\text{11}$-like HOM for a realistic 5-tube nested HC-ARF geometry subjected to random design variations for the tube thicknesses and the tube angle offsets (see Fig. \ref{fig:fig_1}d). Higher-order-mode extinction ratio (HOMER) shown in (c) with exemplary mode-field profiles shown in (d). Mean outer [nested] tube thickness $\overline{t_o}=\SI{393}{\nano\meter}$ [$\overline{t_n}=1.1\cdot\overline{t_o}=\SI{432}{\nano\meter}$]; mean angle offset $\overline{\alpha}=\SI{0}{\degree}$ leading to a mean tube gap separation $\overline{g}=\SI{5.25}{\micro\meter}$;
  Standard dev. of nested tube thickness is 10$\%$ increased with respect to outer tubes, ${\sigma}_{t_n}=1.1\cdot{\sigma}_{t_o}$. The 2D surface plot is generated using a sweep quantization of $\Delta {\sigma}_{t_o}=\SI{1}{\nano\meter}$ and $\Delta {\sigma}_{\alpha}=\SI{0.2}{\degree}$ with a sample size of 25. Noise suppression and interpolation has been applied using a dual-axis regression smoothing algorithm. Contour lines are displayed in a green, dotted shape. Additional design parameters: Core diameter $D_\text{c} = \SI{35}{\micro\meter}$; nested/outer tube diameter ratio $d/D=0.5$. All simulations are performed at wavelength $\lambda=\SI{1.55}{\micro\meter}$.}\label{fig:fig_3}
  \end{center}
\end{figure}

In the first step, we investigated the susceptibility of propagation loss to the effect of random tube thickness variations and random tube angle offsets. 
We apply those effects both independently and simultaneously. Moreover, we also evaluated those effects on HOM propagation and restricted our analyses to $LP_\text{11}$-mode. We, therefore, scan the standard deviation of the outer tube thicknesses from \SIrange[range-units=single]{1}{10}{\micro\meter} [nested: $+10\%$], and of the tube angle offsets from \SIrange{0.2}{2}{\degree}. For every of these combinations, sample groups of 25 random geometries will be generated and evaluated for propagation loss, whereas the fundamental-mode and best propagating HOM were separated into two datasets. The median loss value, determined for each combination, is displayed in a color-coded 2D plot, whereas the color indicates the relative loss increase in $\%$ with respect to the ideal geometry found in subsection \ref{chapter:geom_optim} (here: origin point). The result is displayed in Fig. \ref{fig:fig_3}. Dotted green lines represent the contour of the propagation loss increase. 

We noticed that for the given sample size of 25 values, the corresponding medians can vary erratically, over- and undershooting their neighbors even five quantization units apart. Increasing the sample size to limit this effect to a given threshold is not a reasonable option since computation time already exceeds multiple days for this simulation. Instead, we have chosen to post-process the noisy data by applying an iterative smoothing algorithm. In short words, this custom algorithm interprets the data as a spatially ordered 2D matrix with respect to both axes and then carries out a row- and column-wise polynomial regression of rank 2. The predicted values of the dual-axis regression each produce another matrix with an identical structure to the original, which are then averaged, producing a single intermediate matrix with smoothed up data. This process can be applied repetitively until a certain degree of smoothing is reached. The resulting data converges with rising iteration count. Finally, data interpolation can easily be achieved by evaluating and averaging the final row- and column-wise regressions for any intermediate points necessary. The data shown in Fig. \ref{fig:fig_3}(a,b) has been post-processed utilizing three iterations and interpolating by a factor of 32, effectively eliminating any noise.

For the (a) fundamental mode, the loss increases with either standard deviation rising, reaching a maximum point of $+5.2\%$ at the maximum joint random structural deviation. It can be seen that the effect of random tube angle offsets has a much more pronounced impact on the loss than the varying tube thicknesses. The fairly vertical contour lines confirm this observation, indicating a weak to none impact off the random tube thickness effect for a fixed angle offset. Interestingly, the loss contribution of the angle offset effect is fairly low in the left half of the figure and rapidly picks up in the right half. This is important since it essentially means that angle offset standard deviations up until $\approx$\SI{1}{\degree} do not result in any considerable loss increase $>1\%$, but multiply their impact if this threshold is passed. Circling back to the optimization of the ideal structure, this minor impact is consistent with the general insensitivity of the transmission loss of the five-tube structure to both considered geometrical parameters, as shown in  Fig. \ref{fig:fig_2}a in subsection \ref{chapter:geom_optim}. The HOM loss increase shows an even more extreme domination of the angle offset effect over the tube thickness effect as seen in (b), with a maximum joint loss increase of $65\%$. However, its impact rises more linearly compared to FM. The higher-order-mode extinction ratio (HOMER), which is defined as the ratio between minimum HOM propagation loss and FM propagation loss, indicating the single-modeness of the fiber, is displayed in (c) as a function of both imperfection standard deviations. It can be seen that HOMER solely depends on the angle offset standard deviation and increases approximately proportionally with it. In other words, HOMs are more susceptible to random structural variations than FM. This increased effective single-modeness of perturbed fibers can be interpreted as one of the only advantages of random structural variations when single-mode propagation is desired. In summary, the five-tube nested HC-ARF geometry exhibits negligible susceptibility for random tube thickness variations in the analyzed range and provides a stable FM propagation loss for a moderate random tube angle offset effect. 

\subsection{Wavelength dependency of the imperfections' impact}
\begin{figure}[t!]
  \begin{center}
  \includegraphics[width=3.4in]{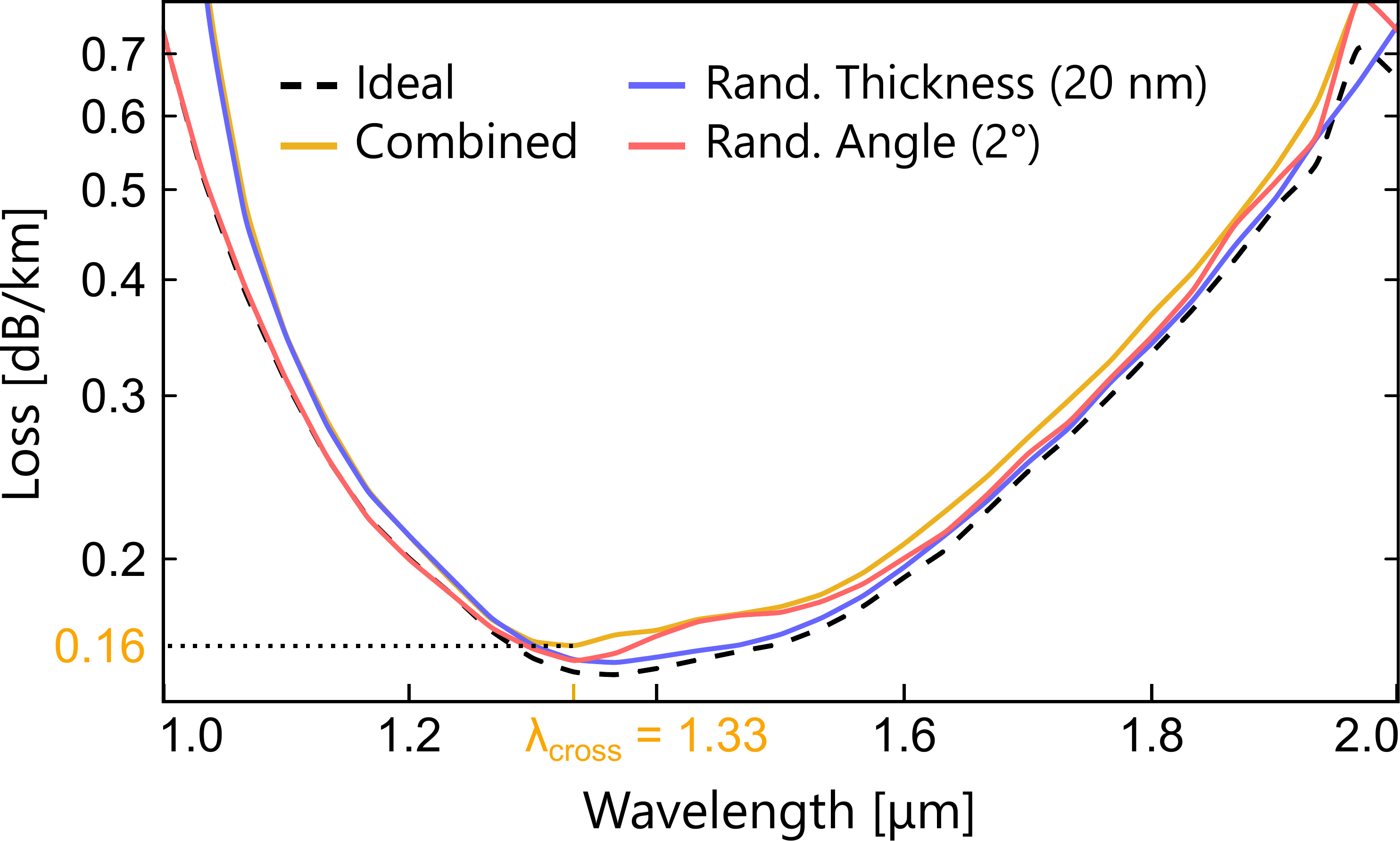}\\
  \caption{Calculated mean propagation loss of $\text{LP}_\text{01}$-like FM for a 5-tube nested HC-ARF affected by different random structural variations as a function of wavelength: Ideal structure (black, dotted); with tube thickness standard dev. ${\sigma}_{t_o}=\SI{20}{\nano\meter}$/${\sigma}_{t_n}=\SI{22}{\nano\meter}$ (blue); with angle offset standard dev. ${\sigma}_{\alpha}=\SI{2}{\degree}$ (red); combined (yellow). A sample size of 10 is used for each category. The sweep quantization is $\Delta {\lambda}=\SI{0.33}{\micro\meter}$. Additional design parameters: Core diameter $D_\text{c} = \SI{35}{\micro\meter}$; nested/outer tube diameter ratio $d/D=0.5$. Mean outer [nested] tube thickness $\overline{t_o}=\SI{393}{\nano\meter}$ [$\overline{t_n}=1.1\cdot\overline{t_o}=\SI{432}{\nano\meter}$]; mean angle offset $\overline{\alpha}=\SI{0}{\degree}$ leading to a mean tube gap separation $\overline{g}=\SI{5.25}{\micro\meter}$.}\label{fig:fig_4}
  \end{center}
\end{figure}
In this subsection, we analyze the same effects, i.e., random tube wall thicknesses as well as random tube angle offsets as a function of the wavelength. Firstly, we want to see how a fiber with realistic fabrication perturbations performs over a broader spectrum, and secondly, we want to identify the random effect with the dominant loss contribution as a function of the wavelength. Therefore, we generate three geometry groups with either a random tube thickness imperfection of ${\sigma}_{t_o}=\SI{20}{\nano\meter}$ [nested +10$\%$], a random angle offset imperfection of ${\sigma}_{\alpha}=\SI{2}{\degree}$, and a simultaneous combination of both. We evaluate those geometries over a wavelength range of \SIrange[range-units=single]{1}{2}{\micro\meter} using a quantization of $\Delta {\lambda}=\SI[number-math-rm = \mathnormal, parse-numbers = false]{0.0\overline{3}}{\micro\meter}$ and plot the respective medians of each group as a function of the wavelength. The result is displayed in Fig. \ref{fig:fig_4}. As expected, the ideal geometry exhibits the lowest propagation loss over the whole spectrum with a minimum value of $\SI{0.15}{dB/km}$ at $\lambda=\SI{1.36}{\micro\meter}$. Both independently applied random effects exhibit a slightly higher loss with the joint configuration forming an upper envelope. Interestingly, the dominating effect switches in the analyzed range at $\lambda_{\text{cross}}=\SI{1.33}{\micro\meter}$. For wavelengths $<\lambda_{\text{cross}}$ the random tube thickness imperfection contributes nearly 100$\%$ of the loss increase. On the contrary, the random angle offset effect poses the main loss contribution in the range between \SIrange[range-units=single]{1.33}{1.65}{\micro\meter}, which is consistent with the previous observations. For higher $\lambda$ values, the shares of both effects nearly balance out with the angle offset effect posing a slightly bigger contribution. It is worth noting that the absolute loss of the fiber category with combined imperfections exhibits a global minimum of \SI{0.16}{dB/km} at the crossing point, which indicates an optimal operation point from the perspective of resilience to random structural perturbations. In conclusion, our simulations show that the shares of the loss contribution and the relative loss increase are highly dependent on the operating wavelength in the light spectrum.

\subsection{Impact of random design variations under a bend condition}

\begin{figure}[t!]
  \begin{center}
  \includegraphics[width=3.4in]{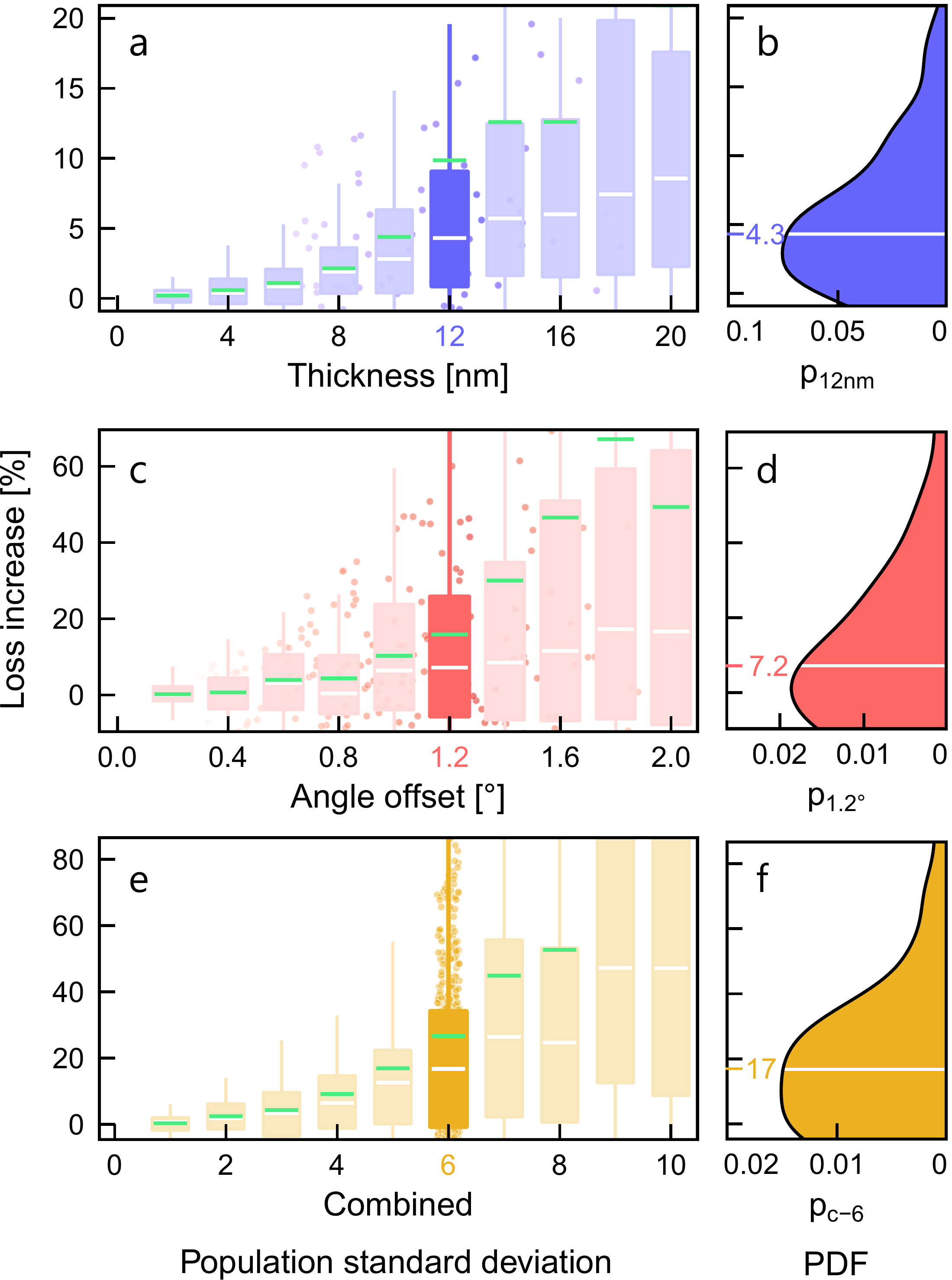}\\
  \caption{Calculated relative transmission loss increase with respect to the ideal structure for imperfect 5-tube nested HC-ARF geometries as a function of the population standard dev. at a constant bend radius of $\SI{5}{\centi\meter}$, visualized as a box plot. (a) Thickness standard dev. scanned from \SIrange[range-units=single]{2}{20}{\nano\meter} [nested +10$\%$], (c) angle offset standard dev. scanned from \SIrange{0.2}{2}{\degree}, (e) imperfections from (a,c) applied simultaneously. Right-sided graphs display estimated probability density functions (PDF) for corresponding highlighted sample groups regarding thickness (b), angle offset (d), and combined (f) structural imperfections. The individual sample groups' median and mean loss increases are indicated by white and green markers, respectively. Individual loss values in the highlighted sample groups of (a) and (c) are plotted over the sample standard dev. in the background. Additional design parameters: Core diameter $D_\text{c} = \SI{35}{\micro\meter}$; nested/outer tube diameter ratio $d/D=0.5$. Mean outer [nested] tube thickness $\overline{t_o}=\SI{393}{\nano\meter}$ [$\overline{t_n}=1.1\cdot\overline{t_o}=\SI{432}{\nano\meter}$]; mean angle offset $\overline{\alpha}=\SI{0}{\degree}$ leading to a mean tube gap separation $\overline{g}=\SI{5.25}{\micro\meter}$. All simulations are performed at wavelength $\lambda=\SI{1.55}{\micro\meter}$.
  }\label{fig:fig_5}
  \end{center}
\end{figure}

In this subsection, we thoroughly analyze and discuss the susceptibility of the fiber propagation loss to the previously described random imperfections while bending the fiber at a constant radius of $\SI{5}{\centi\meter}$ at first and allowing multiple bend radii afterwards. As before, we have studied both effects independently and jointly. Similarly to subsection \ref{section:impact_random_design}, we investigate the impact of different tube thickness standard deviations as well as multiple angle offset standard deviations in the range of \SIrange[range-units=single]{2}{20}{\nano\meter} [nested +10$\%$] and \SIrange{0.2}{2}{\degree}, respectively, with sample groups of size 100. A box plot, grouped by the generating standard deviations, is displayed in Fig. \ref{fig:fig_5}. The independently calculated results are visualized in (a) and (c), and the joint analysis, for which both imperfections are applied simultaneously with the chosen standard deviations, is shown in (e). The median and mean values of the individual sample groups are indicated by white and green markers, respectively. The results are given in relative loss increase to the ideal, bent structure ($R_\text{bend}=\SI{5}{\centi\meter}$), which exhibits a propagation loss of \SI{5.47}{dB/km}. All three graphs show a clear dependence of the loss increase on the standard deviation. The median and mean values are assumed to rise monotonically in the investigated range, whereas the median rises to $8\%$ and $17\%$ for the random tube thickness and angle offset effects, respectively. The maximum median value of $\approx50\%$ for the joint case indicates a mutual amplification in loss increase when both effects are applied concurrently. Estimations of the probability density functions (PDF) for the highlighted sample groups are provided in the corresponding graphs (b,d,f) on the right-hand side. It is evident that the individual geometries exhibit quite a spread of loss values ranging from a multiple of the median loss increase to slightly lower than the propagation loss of the ideal (bent) geometry. To further visualize this spread, the loss increase values corresponding to the geometries inside the highlighted sample groups are plotted over their sample standard deviation for (a) and (c). Compared to Fig. \ref{fig:fig_3}, when viewed independently, the random tube thickness and angle offset imperfections have a more pronounced effect by a factor of approximately 4 and 5 at the end of the scale, respectively. A direct comparison between the joint cases is not possible because of different scales of the thickness standard deviation, but we estimate an increase by a factor of $\approx$7. Concluding, the impact of the investigated random imperfections on five-tube nested HC-ARFs is much more pronounced under a bend condition and can result in a multiple of the relative loss increase compared to that of a straight fiber.

\begin{figure}[b!]
  \begin{center}
  \includegraphics[width=3.4in]{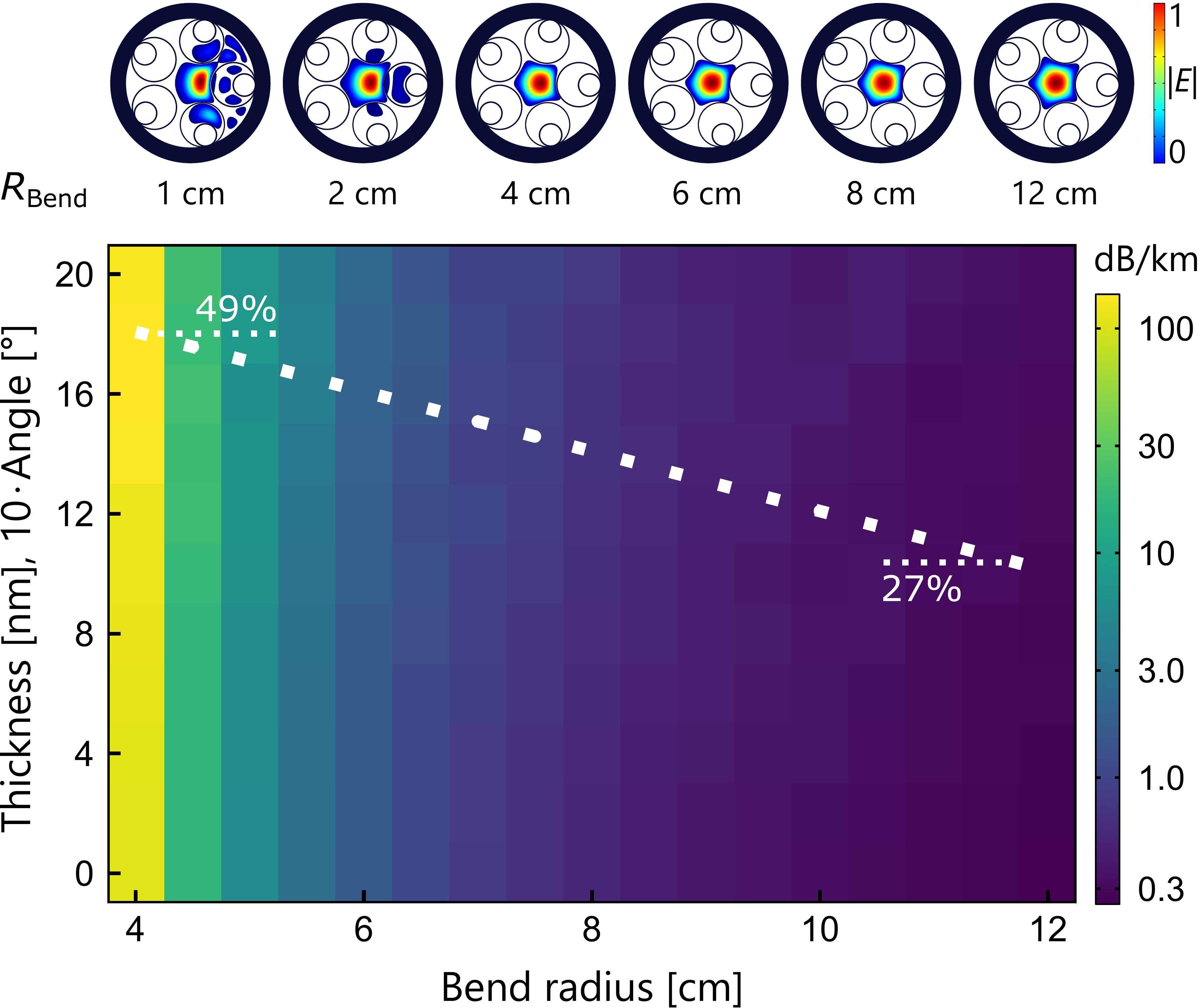}\\
  \caption{Calculated median propagation loss of $\text{LP}_\text{01}$-like FM for a realistic 5-tube nested HC-ARF geometry as a function of the bend radius and the combined random structural variations, namely tube thickness and tube angle offset. The dotted white line denotes the column-wise loss contribution at $\sigma_{t[\alpha]}=\SI{20}{\nano\meter}$ $[\SI{2}{\degree}]$ due to random imperfections. Exemplary mode-field profiles of an imperfect sample geometry as a function of the bend radius are shown at the top. Mean outer [nested] tube thickness $\overline{t_o}=\SI{393}{\nano\meter}$ [$\overline{t_n}=1.1\cdot\overline{t_o}=\SI{432}{\nano\meter}$]; mean angle offset $\overline{\alpha}=\SI{0}{\degree}$ leading to a mean tube gap separation $\overline{g}=\SI{5.25}{\micro\meter}$. The 2D surface plot is generated using a sweep quantization of $\Delta{}r=\SI{0.5}{\centi\meter}$ and $\Delta {\sigma}_{t_o}=\SI{2}{\nano\meter}$, ${\sigma}_{\alpha}=\SI{0.2}{\degree}$ with a sample size of 25. Standard dev. of nested tube thickness is 10$\%$ increased with respect to outer tubes, ${\sigma}_{t_n}=1.1\cdot{\sigma}_{t_o}$. Additional design parameters: Core diameter $D_\text{c} = \SI{35}{\micro\meter}$; nested/outer tube diameter ratio $d/D=0.5$. All simulations are performed at wavelength $\lambda=\SI{1.55}{\micro\meter}$.}\label{fig:fig_6}
  \end{center}
\end{figure}

Bending HC-ARFs to sufficiently small radii in the centimeter scale can result in a tremendous loss increase with respect to the straight configuration\cite{poletti2014nested}. As described in the simulations before, bending a five-tube HC-ARF to a radius of \SI{5}{\centi\meter} increases the absolute propagation loss from \SI{0.17}{dB/km} to \SI{5.47}{dB/km}. We have also shown that bending amplifies the impact of random structural imperfections on the loss. This begs the question how this dynamic behaves over different bend radii and if random imperfections even have to be considered for minimal bend radii. To answer this, we carried out a 2D sweep over practical bend radii in the range of \SIrange[range-units=single]{4}{12}{\centi\meter} while applying a joint imperfection of random tube thicknesses and tube angle offsets with standard deviations ranging from ${\sigma}_t=$ \SIrange[range-units=single]{0}{20}{\nano\meter} and ${\sigma}_{\alpha}=$ \SIrange{0}{2}{\degree}, respectively. A sample group size of 20 has proven to be sufficient. Fig. \ref{fig:fig_6} displays the absolute loss in dB/km with yellow color denoting high loss and dark-blue color denoting low loss. Logarithmic color scaling is used because of the high dynamic range of three orders of magnitude. The resulting median loss ranges between \SIrange{0.28}{140}{dB/km}, with high losses occurring at very low bend radii. At first glance, the variation in the vertical direction, denoting the share of loss contribution due to random design variations, seems very small, if not negligible. However, this is a side effect of the logarithmic scaling. This factor, representing the loss increase from $\sigma=0$ to $\sigma_{t[\alpha]}=\SI{20}{\nano\meter}$ $[\SI{2}{\degree}]$ as a function of the bend radius is displayed by a white, dotted line. It can be seen that the relative loss contribution due to random imperfections does indeed increase with the reciprocal of the bend radius and can lead to a significant worsening of the propagation loss up to $\approx 50\%$ at $R_\text{bend}=\SI{4}{\centi\meter}$. As a rule of thumb, one can assume that the loss contribution due to random perturbations increases by approximately $3\%$ per centimeter less bend radius in the studied range. Exemplary mode-field profiles of an imperfect sample geometry as a function of the bend radius are shown at the top of Fig. \ref{fig:fig_6}. [Little more here, 1-2 sentences or so.]
 
\subsection{Further imperfections and outlook}
The imperfections considered in this paper are non-doubly the most visually pronounced ones, however, other imperfections have been observed which seek further analysis. One of them is an angular offset of the nested tubes, independent of their corresponding surrounding tube. This creates the effect of "rolling" nested tubes inside their surrounding tube within the range of a few degrees. Also, a randomly distributed non-circular, anisotropic shape of the cladding structure can be investigated. The authors suggest deploying machine learning techniques to predict the expected fiber characteristics as an alternative to the classical finite-element approach.
These investigations are left for further research. 
\section{Conclusion}
\label{section:conclusion}
We have investigated the impact of random structural variations on nested hollow-core anti-resonant fibers on the propagation loss using numerical simulations. A five-tube geometry has been selected and optimized with respect to the tube wall thickness and gap separation to serve as an ideal reference point. The random structural effects considered in this article were varying outer and nested tube wall thicknesses as well as tube angle offsets, whereas the second one has been identified to dominate in the majority of tested operating conditions. Analysis of single- and multi-mode propagation loss as a function of the imperfection intensity at a fixed wavelength of $\SI{1.55}{\micro\meter}$ has shown that random angle offsets have a much more pronounced impact on the loss than random tube thicknesses by a factor of 5, leading to an approximate combined loss increase of $5\%$ and $65\%$ for single- and multi-mode propagation, respectively. However, the share is very wavelength-dependent and even reaches a point where the dominant contribution flips to the random tube thickness effect. An interesting phenomenon worth noting is that the fiber with both imperfections applied simultaneously has its lowest (absolute) propagation loss at this flipping point. Focusing only on the fundamental mode, we have also studied how the impact of random structural variations changes when a bend condition is applied. We found that the loss contribution by random effects rises with the reciprocal of the bend radius and has an approximately seven-fold impact at $R_b=\SI{5}{\centi\meter}$ compared to no bend condition. Overall, a significant worsening of the propagation loss of about 50$\%$ due to random structural effects can be expected at a bend radius of \SI{4}{\centi\meter}.

\section*{Acknowledgment}
The authors would like to thank Dr. Rodrigo Amezcua-Correa and Dr. Francesco Poletti for useful discussions.
\ifCLASSOPTIONcaptionsoff
  \newpage
\fi

\bibliography{IEEEabrv,references}

\bibliographystyle{IEEEtran}

%
\begin{IEEEbiography}[{\includegraphics[width=1.05in,height=1.25in,clip,keepaspectratio]{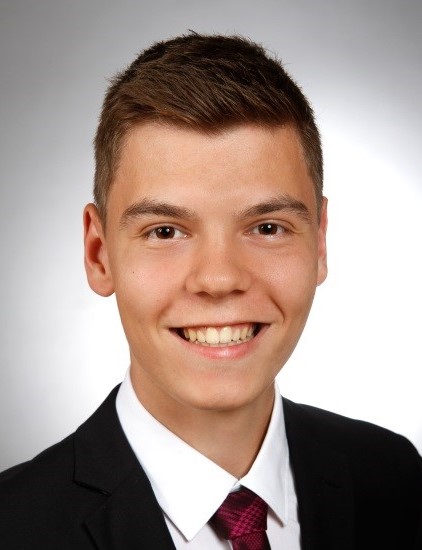}}]{Michael Petry}



(S’21) received his B.Eng. degree (with Distinction) in Electrical Engineering from the Karlsruhe University of Applied Sciences (HKA), Germany in 2021. He is currently pursuing two M.Sc degrees in Electrical Engineering at Florida Polytechnic University and HKA as part of the German-American Fulbright exchange program with expected graduation in May 2022 and February 2023, respectively. As a member of the university's optical fiber group, his research focuses on single- and multi-mode fiber design, statistical characterization and optimization. He is also working on incorporating Machine Learning (ML) techniques into the HC-ARF domain to provided an alternative approach for fiber analysis. Further fields of interest are information theory and coding using Deep Learning (DL), Radio Frequency (RF), Finite-Element (FE) techniques, VLSI Design and secure computer architecture. He is an IEEE student member.
\end{IEEEbiography}

\begin{IEEEbiography}[{\includegraphics[width=1.05in,height=1.25in,clip,keepaspectratio]{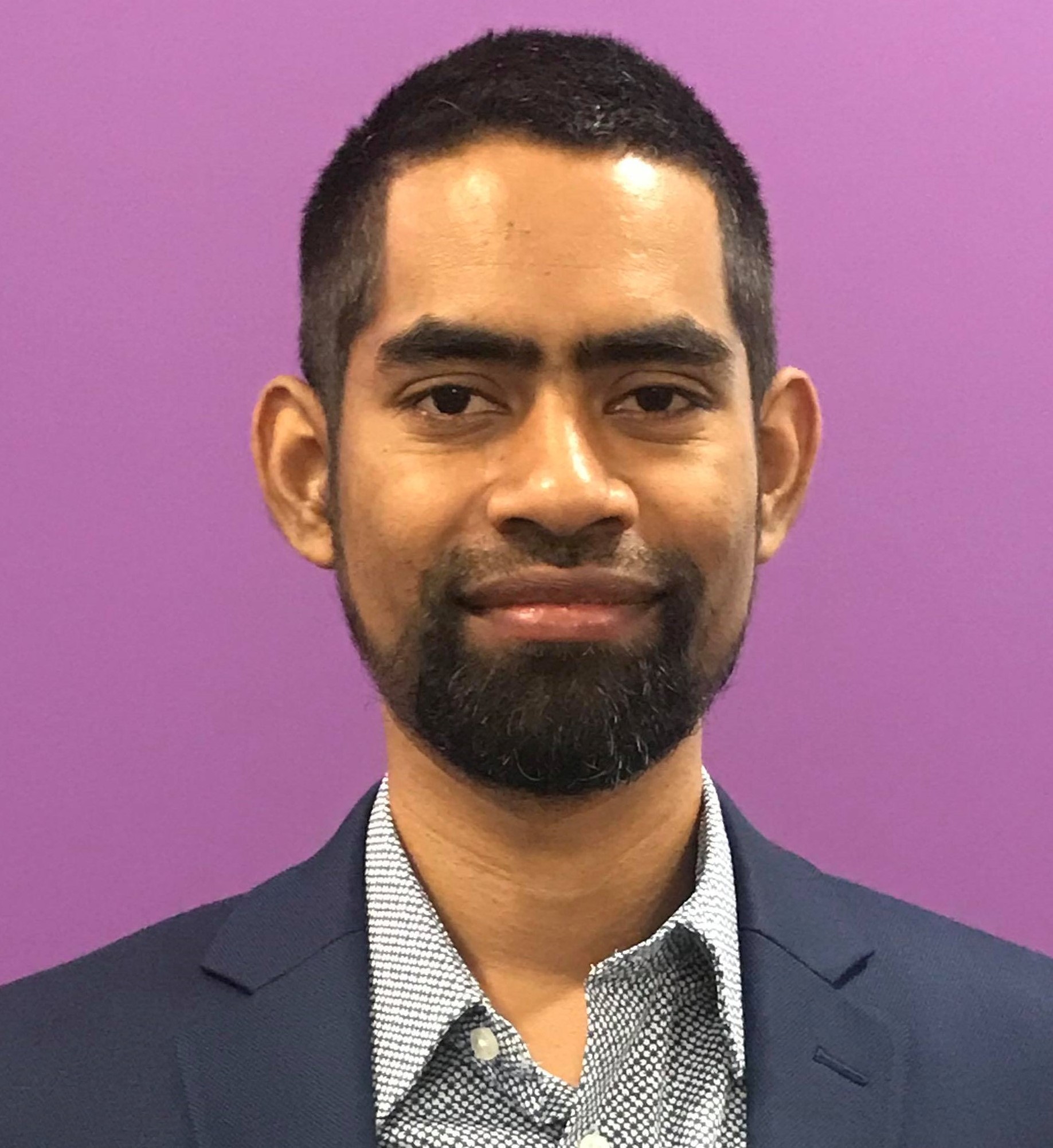}}]{Md. Selim Habib}



(S’13–SM’19) received the B.Sc. and M.Sc. Degrees in Electrical and Electronic Engineering from Rajshahi University of Engineering and Technology, Rajshahi, Bangladesh in 2008 and 2012 respectively. He received the Ph.D. degree in photonics engineering from Technical University of Denmark (DTU) in 2017. After finishing his Ph.D., he joined as a Postdoctoral Researcher in Fibers Sensors and Supercontinuum Group at the Department of Photonics Engineering, DTU. After finishing his Postdoctoral Fellowship at DTU, he worked as a Postdoctoral Research Associate at CREOL, The College of Optics and Photonics, University of Central Florida, USA from September, 2017 to August, 2019. Now he is an Assistant Professor of Electrical and Computer Engineering at Florida Polytechnic University, USA. His research mainly focuses on design, fabrication, and characterization of low loss hollow-core fiber in the near-IR to mid-IR, light gas nonlinear interaction in hollow-core fibers, supercontinuum generation, and multi-mode nonlinear optics. He has published more than 40 articles in referred journals. 

Dr. Habib is a Senior Member of Institute of Electrical and Electronics Engineers (IEEE), Optical Society of America (OSA) Early Careers Member, and Executive officer of OSA Fiber modeling and Fabrication group. Dr. Habib is an Associate Editor of IEEE Access, and Feature Editor of Applied Optics (OSA). He received the University Gold Medal Award from Rajshahi University of Engineering and Technology in 2014.
\end{IEEEbiography}

\vfill

\balance
\end{document}